# Psychological games of entangled players.

Michail Zak
Senior Research Scientist (Emeritus)
Jet Propulsion Laboratory California Institute of Technology
Pasadena, CA 91109

# Abstract.

The paper introduces a new approach to theory of differential games in which entangled players try to predict and influence actions of their adversaries. The entanglement is generated by a joint probability density known by the players. In case of a complex density, its imaginary part represents a measure of uncertainty of the density distribution.  The novelty of the approach is in non-Newtonian mathematical formalism that is based upon a **behavioral** model of Livings. The model is quantum inspired: it represented by a modified Madelung equation in which the quantum potential is replaced by different, specially chosen "computational" potential. It consists of motor dynamics simulating actual behavior of the object, and mental dynamics representing evolution of the corresponding knowledge-base and incorporating this knowledge in the form of information flows into the motor dynamics. Due to feedback from mental dynamics, the motor dynamics attains quantum-like properties: its trajectory splits into a family of different trajectories, and each of those trajectories can be chosen with the probability prescribed by the mental dynamics. The model addresses a new type of entanglement that correlates the *probabilities* of actions of Livings rather than the actions themselves

## 1. Introduction.

This paper is devoted to a new approach to differential games, i.e. to a group of problems related to the modeling and analysis of conflict in the context of a dynamical system. We will concentrate on the non-Newtonian properties of dynamics describing a psychology-based behavior of intelligent agents as players.

***a. Justification for non-Newtonian approach.*** All the previous attempts to develop models for so called active systems (i.e., systems that possess certain degree of autonomy from the environment that allows them to perform motions that are not directly controlled from outside) have been based upon the principles of Newtonian and statistical mechanics. These models appear to be so general that they predict not only physical, but also some biological and economical, as well as social patterns of behavior exploiting such fundamental properties of nonlinear dynamics as attractors. Not withstanding indisputable successes of that approach (neural networks, distributed active systems, etc.) there is still a fundamental limitation that characterizes these models on a dynamical level of description: they propose no difference between a solar system, a swarm of insects, and a stock market. Such a phenomenological reductionism is incompatible with the first principle of progressive biological evolution associated with Darwin. According to this principle, the evolution of living systems is directed toward the highest levels of complexity if the complexity is measured by an irreducible number of different parts, which interact in a well-regulated fashion (although in some particular cases deviations from this general tendency are possible).  At the same time, the solutions to the models based upon dissipative Newtonian dynamics eventually approach attractors where the evolution stops while these attractors dwell on the subspaces of lower dimensionality, and therefore, of the lower complexity (until a "master" reprograms the model). Therefore, such models fail to provide an autonomous progressive evolution of living systems (i.e. evolution leading to increase of complexity). Let us now extend the dynamical picture to include thermal forces. That will correspond to the stochastic extension of Newtonian models, while the Liouville equation will extend to the Fokker-Planck equation that includes thermal force effects through the diffusion term.  Actually, it is a well-established fact that evolution of life has a diffusion-based stochastic nature as a result of the multi-choice character of behavior of living systems. Such an extended thermodynamics-based approach is more relevant to model of living systems, and therefore, the simplest living species must obey the second law of thermodynamics as physical particles do. However, then the evolution of living systems (during periods of their isolation) will be regressive since their entropy will increase. Therefore, Newtonian physics is not sufficient for simulation the specific properties typical for intelligence.

There is another argument in favor of a non-Newtonian approach to modeling intelligence. As pointed out by Penrose, [2] the Gödel's famous theorem has the clear implication that mathematical understanding cannot be reduced to a set of known computational rules. That means that no knowable set of purely computational procedures could lead to a computer-control robot that possesses genuine mathematical understanding. In other words, such privileged properties of intelligent systems as common sense, intuition, or consciousness are non-computable within the framework of classical models. That is why a fundamentally new physics is needed to capture these "mysterious" aspects of intelligence, and in particular, to decision making process.

**2. Dynamical model for simulations**.
In this section we introduce and discuss a **behavioral** model of intelligent agents, or players. The model is based upon departure from Newtonian dynamics to quantum inspired dynamics that was first introduced in [6,7].

***a. Destalizing effect of Liouville feedback.*** We will start with derivation of an auxiliary result that illuminates departure from Newtonian dynamics. For mathematical clarity, we will consider here a one-dimensional motion of a unit mass under action of a force $f$ depending upon the *velocity $v$* and time $t$ and present it in a dimensionless form

$$\dot{v} = f(v,t) \tag{1}$$

referring all the variables to their representative values $v_0, t_0, etc.$

If initial conditions are not deterministic, and their probability density is given in the form

$$\rho_0 = \rho_0(V), \qquad where \quad \rho \geq 0, \quad and \quad \int_{-\infty}^{\infty} \rho dV = 1 \tag{2}$$

while $\rho$ is a *single- valued* function, then the evolution of this density is expressed by the corresponding Liouville equation

$$\frac{\partial \rho}{\partial t} + \frac{\partial}{\partial V}(\rho f) = 0 \tag{3}$$

The solution of this equation subject to initial conditions and normalization constraints (2) determines probability density as a function of $V$ and $t$:

$$\rho = \rho(V,t) \tag{4}$$

*Remark.* Here and below we make distinction between the random *variable v(t)* and its *values V* in probability space.

In order to deal with the constraint (2), let us integrate Eq. (3) over the whole space assuming that $\rho \to 0$ at $|V| \to \infty$ and $|f| < \infty$ . Then

$$\frac{\partial}{\partial t} \int_{-\infty}^{\infty} \rho dV = 0, \int_{-\infty}^{\infty} \rho dV = const, \tag{5}$$

Hence, the constraint (2) is satisfied for $t > 0$ if it is satisfied for $t = 0$.

Let us now specify the force $f$ as a feedback from the Liouville equation

$$f(v,t) = \varphi[\rho(v,t)] \tag{6}$$

and analyze the motion after substituting the force (6) into Eq.(2)

$$\dot{v} = \varphi[\rho(v,t)], \tag{7}$$

This is a fundamental step in our approach. Although the theory of ODE does not impose any restrictions upon the force as a function of space coordinates, the Newtonian physics does: equations of motion are never coupled with the corresponding Liouville equation. Moreover, it can be shown that such a coupling leads to non-Newtonian properties of the underlying model. Indeed, substituting the force $f$ from Eq. (6) into Eq. (3), one arrives at the *nonlinear* equation of evolution of the probability density

$$\frac{\partial \rho}{\partial t} + \frac{\partial}{\partial V}\{\rho \varphi[\rho(V,t)]\} = 0 \tag{8}$$

Let us now demonstrate the destabilizing effect of the feedback (6). For that purpose, it should be noticed that the derivative $\partial \rho / \partial v$ must change its sign at least once, within the interval $-\infty < v < \infty$ , in order to satisfy the normalization constraint (2).

But since

$$Sign \frac{\partial \dot{v}}{\partial v} = Sign \frac{d\varphi}{d\rho} Sign \frac{\partial \rho}{\partial v} \tag{9}$$

there will be regions of *v* where the motion is unstable, and this instability generates randomness with the probability distribution guided by the Liouville equation (8). It should be noticed that the condition (9) may lead to exponential or polynomial growth of *v* (in the last case the motion is called neutrally stable, however, as will be shown below, it causes the emergence of randomness as well if prior to the polynomial growth, the Lipcshitz condition is violated).

***b. Emergence of self-generated stochasticity.*** In order to illustrate mathematical aspects of the concepts of Liouville feedback in systems under consideration as well as associated with it instability and randomness, let us take the feedback (6) in the form

$$f = -\sigma^2 \frac{\partial}{\partial v} \ln \rho, \qquad (10)$$

to obtain the following equation of motion

$$\dot{v} = -\sigma^2 \frac{\partial}{\partial v} \ln \rho, \qquad (11)$$

This equation should be complemented by the corresponding Liouville equation (in this particular case, the Liouville equation takes the form of the Fokker-Planck equation)

$$\frac{\partial \rho}{\partial t} = \sigma^2 \frac{\partial^2 \rho}{\partial V^2} \qquad (12)$$

Here *v* stands for a particle velocity, and $\sigma^2$ is the constant diffusion coefficient.

The solution of Eq. (12) subject to the sharp initial condition

$$\rho = \frac{1}{2\sigma\sqrt{\pi t}} \exp(-\frac{V^2}{4\sigma^2 t}) \qquad (13)$$

describes diffusion of the probability density, and that is why the feedback (10) will be called a diffusion feedback.

Substituting this solution into Eq. (11) at *V=v* one arrives at the differential equation with respect to *v (t)*

$$\dot{v} = \frac{v}{2t} \qquad (14)$$

and therefore,

$$v = C\sqrt{t} \qquad (15)$$

where *C* is an arbitrary constant. Since *v=0* at *t=0* for any value of *C*, the solution (15) is consistent with the sharp initial condition for the solution (13) of the corresponding Liouville equation (12). The solution (15) describes the simplest irreversible motion: it is characterized by the "beginning of time" where all the trajectories intersect (that results from the violation of Lipcsitz condition at *t=0*, Fig.9), while the backward motion obtained by replacement of *t* with *(-t)* leads to imaginary values of velocities. One can notice that the probability density (13) possesses the same properties.

It is easily verifiable that the solution (15) has the same structure as the solution of the Madelung equation [9], although the dynamical system (11), (12) is not quantum! The explanation of such a "coincidence" is very simple: the system (11), (12) has the same dynamical topology as that of the Madelung equation where the equation of conservation of the probability is coupled with the equation of conservation of the momentum. As will be shown below, the system (11), (12) neither quantum nor Newtonian, and we will call such systems quantum-inspired, or self-supervised.

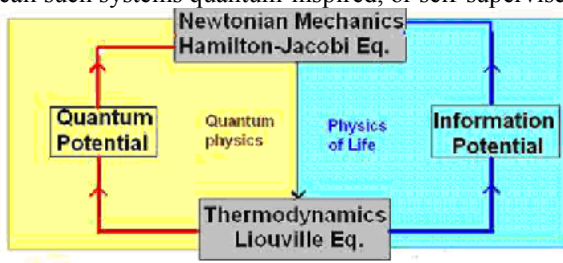

**Figure 1 . Classical Physics, Quantum physics, and Physics of Life**

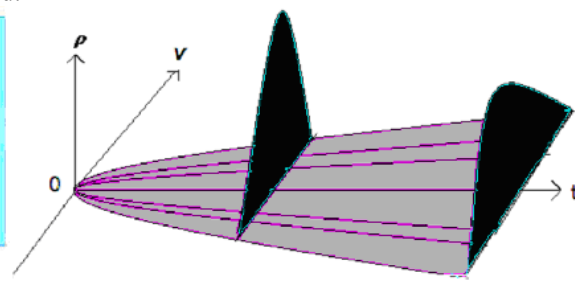

*Figure 2   Stochastic process and probability density*

Further analysis of the solution (15) demonstrates that the solution (15) is *unstable* since

$$\frac{d\dot{v}}{dv} = \frac{1}{2t} > 0 \qquad (16)$$

and therefore, an initial error always grows generating *randomness*. Initially, at *t=0*, this growth is of infinite rate since the Lipchitz condition at this point is violated

$$\frac{d\dot{v}}{dv} \to \infty \ at \qquad t \to 0 \qquad (17)$$

This type of instability has been introduced and analyzed in [4]. The unstable equilibrium point ( $v = 0$ ) has been called a terminal attractor, and the instability triggered by the violation of the Lipchitz condition – a non-Lipchitz instability. The basic property of the non- Lipchitz instability is the following: if the initial condition is infinitely close to the repeller, the transient solution will escape the repeller during a *bounded* time while for a regular repeller the time would be *unbounded*. Indeed, an escape from the simplest regular repeller can be described by the exponent $v = v_0 e^t$. Obviously $v \to 0$ if $v_0 \to 0$, unless the time period is unbounded. On the contrary, the period of escape from the terminal attractor (15) is bounded (and even infinitesimal) if the initial condition is infinitely small, (see Eq. (17)).

Considering first Eq. (15) at fixed *C* as a sample of the underlying stochastic process (13), and then varying C, one arrives at the whole ensemble characterizing that process, (see Fig. 2). One can verify that, as follows from Eq. (13), [3], the expectation and the variance of this process are, respectively

$$\bar{v} = 0, \quad \tilde{v} = 2\sigma^2 t \qquad (18)$$

The same results follow from the ensemble (15) at $-\infty \le C \le \infty$. Indeed, the first equality in (18) results from symmetry of the ensemble with respect to *v=0*; the second one follows from the fact that

$$\tilde{v} \propto v^2 \propto t \qquad (19)$$

It is interesting to notice that the stochastic process (15) is an alternative to the following Langevin equation, [3]

$$\dot{v} = \Gamma(t), \quad \bar{\Gamma} = 0, \quad \tilde{\Gamma} = \sigma \qquad (20)$$

that corresponds to the *same* Fokker-Planck equation (12). Here $\Gamma(t)$ is the Langevin (random) force with zero mean and constant variance $\sigma$.

Thus, the emergence of self-generated stochasticity is the first basic non-Newtonian property of the dynamics with the Liouville feedback.

***c. Second law of thermodynamics***. In order to demonstrate another non-Newtonian property of the systems considered above, let us start with the dimensionless form of the Langevin equation for a one-dimensional Brownian motion of a particle subjected to a random force, [3]

$$\dot{v} = -kv + \Gamma(t), \quad < \Gamma(t) >= 0, \quad < \Gamma(t)\Gamma(t') >= 2\sigma\delta(t-t') , \ [\Gamma] = 1/s \qquad (21)$$

Here $v$ is the dimensionless velocity of the particle (referred to a representative velocity $v_0$ ), $k$ is the coefficient of a linear damping force, $\Gamma(t)$ is the Langevin (random) force per unit mass, $\sigma > 0$ is the noise strength. The representative velocity $v_0$ can be chosen, for instance, as the initial velocity of the motion under consideration.

The corresponding continuity equation for the probability density $\rho$ is the following Fokker-Planck equation

$$\frac{\partial \rho}{\partial t} = k\frac{\partial (V\rho)}{\partial V} + \sigma\frac{\partial^2 \rho}{\partial V^2}, \ \int_{-\infty}^{\infty} \rho dV = 1 \qquad (22)$$

Obviously without external control, the particle cannot escape from the Brownian motion.

Let us now introduce a new force (referred to unit mass and divided by $v_0$ ) as a Liouville feedback

$$f = \sigma \exp\sqrt{D}\ \frac{\partial}{\partial v}\ln\rho, \qquad [f] = 1/s \qquad (23)$$

Here $D$ is the dimensionless variance of the stochastic process $D(t) = \int_{-\infty}^{\infty} \rho V^2 dV$ ,

Then the new equation of motion takes the form

$$\dot{v} = -kv + \Gamma(t) + \sigma \exp \sqrt{D} \frac{\partial}{\partial v} \ln \rho, \tag{24}$$

and the corresponding Fokker-Planck equation becomes nonlinear

$$\frac{\partial \rho}{\partial t} = k \frac{\partial (V\rho)}{\partial V} + \sigma(1 - \exp \sqrt{D}) \frac{\partial^2 \rho}{\partial V^2}, \qquad \int_{-\infty}^{\infty} \rho dV = 1 \tag{25}$$

Obviously the diffusion coefficient in Eq. (25) is *negative*. Multiplying Eq.(25) by $V^2$, then integrating it with respect to $V$ over the whole space, one arrives at ODE for the variance $\widetilde{v}$ $(t)$

$$\dot{D} = 2[\sigma(1 - \exp \sqrt{D}) - kD] \tag{26}$$

Thus, as a result of *negative* diffusion, the variance $D$ monotonously vanishes regardless of the initial value $D$ *(0)*. It is interesting to note that the time $T$ of approaching $D =0$ is finite

$$T = \frac{1}{2} \int_{D(0)}^{0} \frac{dD}{\sigma(1 - \exp \sqrt{D}) - kD} \le \frac{1}{2\sigma} \int_{0}^{\infty} \frac{dD}{\exp \sqrt{D} - 1} = \frac{\pi}{6\sigma} \tag{27}$$

This terminal effect is due to violation of the Lipchitz condition, at $D = 0$, [4].

Let us review the structure of the force (23): it is composed only out of the probability density and its variance, i.e. out of the components of the conservation equation (25); at the same time, Eq. (25) itself is generated by the equation of motion (24). Consequently, the force (23) is *not* an external force. Nevertheless, it allows the particle escape from the Brownian motion using its own "internal effort". It would be reasonable to call the force (23) an *information force* since it links to information rather than to energy.

Thus, we came across the phenomenon that violates the second law of thermodynamics when the dynamical system moves from disorder to the order without external interactions due to a feedback from the equation of conservation of the probability to the equation of conservation of the momentum One may ask why the negative diffusion was chosen to be nonlinear. Let us turn to a linear version of Eq. (26)

$$\frac{\partial \rho}{\partial t} = -\sigma^2 \frac{\partial^2 \rho}{\partial V^2}, \quad \int_{-\infty}^{\infty} \rho dV = 1 \tag{28}$$

and discuss a negative diffusion in more details. As follows from the linear equivalent of Eq. (26)

$$\dot{D} = -2\sigma, i.e. \qquad D = D_0 - 2\sigma t < 0 \ at \qquad t > D_0 / (2\sigma) \tag{29}$$

Thus, eventually the variance becomes negative, and that disqualifies Eq. (29) from being meaningful. As shown in [6], the initial value problem for this equation is ill-posed: its solution is not differentiable at any point. Therefore, a *negative diffusion must be nonlinear* in order to protect the variance from becoming negative, Fig.3.

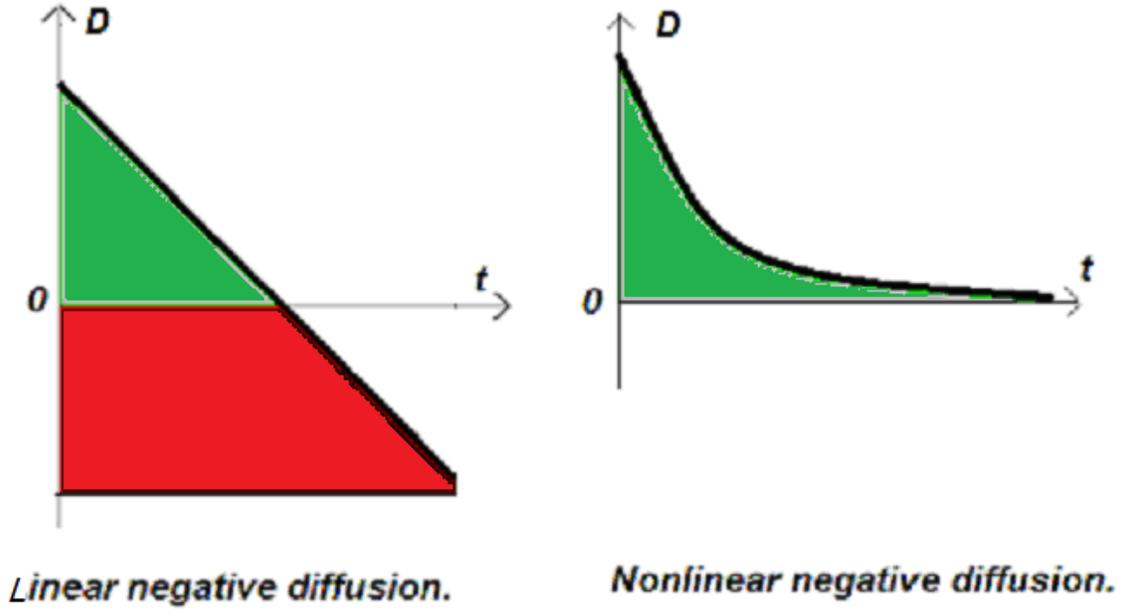

*Linear negative diffusion.*       *Nonlinear negative diffusion.*

**Figure 3. Negative diffusion.**

It should be emphasized that negative diffusion represents a major departure from both Newtonian mechanics and classical thermodynamics by providing a progressive evolution of complexity against the Second Law of thermodynamics.

In the next sub-section we will demonstrate again that formally the dynamics introduced above does not belong to the Newtonian world; nevertheless its self-supervising capability may associate such a dynamics with a potential model for intelligent behavior. For that purpose we will turn to even simpler version of this dynamics by removing the external Langevin force and simplifying the information force.

In 1945 Schrödinger wrote in his book "What is life": *"Life is to create order in the disordered environment against the second law of thermodynamics".* The self-supervised dynamical system introduced above is fully consistent with this statement. Indeed, consider a simplified version of Eqs. (21) and (22)

$$\dot{v} = \sigma \sqrt{D} \frac{\partial}{\partial v} \ln \rho, \qquad (30)$$

$$\frac{\partial \rho}{\partial t} = -\sigma \sqrt{D} \frac{\partial^2 \rho}{\partial V^2}, \qquad \int_{-\infty}^{\infty} \rho dV = 1 \qquad (31)$$

Removal of the Langevin forces makes the particle *isolated*. Nevertheless the particle has a capability of moving from disorder to order. For demonstration of this property we will assume that the Langevin force was suddenly removed at $t = 0$ so that the initial variance $D_0 > 0$. Then

$$\dot{D} = -2\sigma \sqrt{D} \qquad (32)$$

whence $\qquad D = (\sqrt{D_0} - \sigma t)^2 \qquad (33)$

As follows from Eq. (33), as a result of *internal, self-generated* force

$$F = \sigma \sqrt{D} \frac{\partial}{\partial v} \ln \rho, \qquad (34)$$

the Brownian motion gradually disappears and then vanishes abruptly:

$$D \to 0, \qquad \dot{D} \to 0, \qquad \frac{d\dot{D}}{dD} \to \infty \quad at \qquad t \to \frac{\sqrt{D_0}}{\sigma} \qquad\qquad (35)$$

Thus the probability density shrinks to a delta-function at $t = \dfrac{\sqrt{D_0}}{\sigma}$. Consequently, the entropy

$H(t) = -\displaystyle\int_V \rho \ln \rho \, dV$ decreases down to zero, and that violates the second law of thermodynamics.

Another non-Newtonian property is entanglement.

***d. Entanglement.*** In this sub-section we will introduce a fundamental and still mysterious property that was predicted theoretically and corroborated experimentally in quantum systems: entanglement. Quantum entanglement is a phenomenon in which the quantum states of two or more objects have to be described with reference to each other, even though the individual objects may be spatially separated. This leads to correlations between observable physical properties of the systems. As a result, measurements performed on one system seem to be instantaneously influencing other systems entangled with it. Different views of what is actually occurring in the process of quantum entanglement give rise to different interpretations of quantum mechanics. Here we will demonstrate that entanglement is not a prerogative of quantum systems: it occurs in *quantum-inspired (QI)* systems that are under consideration in this paper.. That will shed light on the concept of entanglement as a special type of global constraint imposed upon a broad class of dynamical systems that includes quantum as well as quantum-inspired (*QI*) systems.

In order to introduce entanglement in *QI* system, we will start with Eqs.(11) and (12) and generalize them to the two-dimensional case

$$\dot{v}_1 = -a_{11} \frac{\partial}{\partial v_1} \ln \rho - a_{12} \frac{\partial}{\partial v_2} \ln \rho, \qquad\qquad (36)$$

$$\dot{v}_2 = -a_{21} \frac{\partial}{\partial v_1} \ln \rho - a_{22} \frac{\partial}{\partial v_2} \ln \rho, \qquad\qquad (37)$$

$$\frac{\partial \rho}{\partial t} = a_{11} \frac{\partial^2 \rho}{\partial V^2} + (a_{12} + a_{21}) \frac{\partial^2 \rho}{\partial V_1 \partial V_2} + a_{22} \frac{\partial^2 \rho}{\partial V_2^2}, \qquad\qquad (38)$$

As in the one- dimensional case, this system describes diffusion without a drift

The solution to Eq. (38) has a closed form

$$\rho = \frac{1}{\sqrt{2\pi \det[\hat{a}_{ij}]t}} \exp(-\frac{1}{4t} b'_{ij} V_i V_j), \, i = 1,2. \qquad\qquad (39)$$

Here

$$[b'_{ij}] = [\hat{a}_{ij}]^{-1} \, , \hat{a}_{11} = a_{11}, \hat{a}_{22} = a_{22}, \hat{a}_{12} = \hat{a}_{21} = a_{12} + a_{21}, \; \hat{a}_{ij} = \hat{a}_{ji}, b'_{ij} = b'_{ji}, \qquad (40)$$

Substituting the solution (39) into Eqs. (36) and (37), one obtains

$$\dot{v}_1 = \frac{b_{11} v_1 + b_{12} v_2}{2t} \qquad\qquad (41)$$

$$\dot{v}_2 = \frac{b_{21} v_1 + b_{22} v_2}{2t}, \qquad\qquad b_{ij} = b'_{ij} \hat{a}_{ij} \qquad\qquad (42)$$

Eliminating $t$ from these equations, one arrives at an ODE in the configuration space

$$\frac{dv_2}{dv_1} = \frac{b_{21} v_1 + b_{22} v_2}{b_{11} v_1 + b_{12} v_2}, \qquad\qquad v_2 \to 0 \quad at \qquad v_1 \to 0, \qquad (43)$$

This is a classical singular point treated in text books on ODE.

Its solution depends upon the roots of the characteristic equation

$$\lambda^2 - 2b_{12}\lambda + b^2{}_{12} - b_{11} b_{22} = 0 \qquad\qquad (44)$$

Since both the roots are real in our case, let us assume for concreteness that they are of the same sign, for instance, $\lambda_1 = 1, \quad \lambda_2 = 1$. Then the solution to Eq. (43) is represented by the family of straight lines

$$v_2 = \tilde{C} v_1, \qquad\qquad \tilde{C} = const.$$

Substituting this solution into Eq. (41) yields $\qquad\qquad\qquad\qquad (45)$

$$v_1 = Ct^{\frac{1}{2}(b_{11} + \widetilde{C}b_{12})} \qquad\qquad v_2 = \widetilde{C}Ct^{\frac{1}{2}(b_{11} + \widetilde{C}b_{12})} \tag{46}$$

Thus, the solutions to Eqs. (36) and (37) are represented by two-parametrical families of random samples, as expected, while the randomness enters through the time-independent parameters $C$ and $\widetilde{C}$ that can take any real numbers. Let us now find such a combination of the variables that is deterministic. Obviously, such a combination should not include the random parameters $C$ or $\widetilde{C}$. It easily verifiable that

$$\frac{d}{dt}(\ln v_1) = \frac{d}{dt}(\ln v_2) = \frac{b_{11} + \widetilde{C}b_{12}}{2t} \tag{47}$$

and therefore,

$$(\frac{d}{dt}\ln v_1)/(\frac{d}{dt}\ln v_2) \equiv 1 \tag{48} \qquad\qquad \text{(II.1.2.61)}$$

Thus, the ratio (48) is deterministic although both the numerator and denominator are random,(see Eq.(47)). This is a fundamental non-classical effect representing a global constraint. Indeed, in theory of stochastic processes, two random functions are considered statistically equal if they have the same statistical invariants, but their point-to-point equalities are not required (although it can happen with a vanishingly small probability). As demonstrated above, the *diversion of determinism into randomness via instability (due to a Liouville feedback), and then conversion of randomness to partial determinism (or coordinated randomness) via entanglement* is the fundamental non-classical paradigm that may lead to instantaneous transmission of *conditional* information on remote distance that to be discussed below

*e. Relevance to model of intelligent agents*. The model under discussion was inspired by E. Schrödinger, the creator of quantum mechanics who wrote in his book "What is Life": "Life is to create order in the disordered environment against the second law of thermodynamics". The proposed model illuminates the "border line" between living and non-living systems. The model introduces a biological particle that, in addition to Newtonian properties, possesses the ability to process information. The probability density can be associated with the *self-image* of the biological particle as a member of the class to which this particle belongs, while its ability to convert the density into the information force - with the *self-awareness* (both these concepts are adopted from psychology). Continuing this line of associations, the equation of motion (such as Eqs (1) or (7)) can be identified with a motor dynamics, while the evolution of density (see Eqs. (3) or (8) –with a mental dynamics. Actually the mental dynamics plays the role of the Maxwell sorting demon: it rearranges the probability distribution by creating the information potential and converting it into a force that is applied to the particle. One should notice that mental dynamics describes evolution of the whole class of state variables (differed from each other only by initial conditions), and that can be associated with the ability to generalize that is a privilege of living systems. Continuing our biologically inspired interpretation, it should be recalled that the second law of thermodynamics states that the entropy of an isolated system can only increase. This law has a clear probabilistic interpretation: increase of entropy corresponds to the passage of the system from less probable to more probable states, while the highest probability of the most disordered state (that is the state with the highest entropy) follows from a simple combinatorial analysis. However, this statement is correct only if there is no Maxwell' sorting demon, i.e., nobody inside the system is rearranging the probability distributions. But this is precisely what the Liouville feedback is doing: it takes the probability density $\rho$ from Equation (3), creates functionals and functions of this density, converts them into a force and applies this force to the equation of motion (1). As already mentioned above, because of that property of the model, the evolution of the probability density becomes nonlinear, and the entropy may decrease "against the second law of thermodynamics", Fig.4.

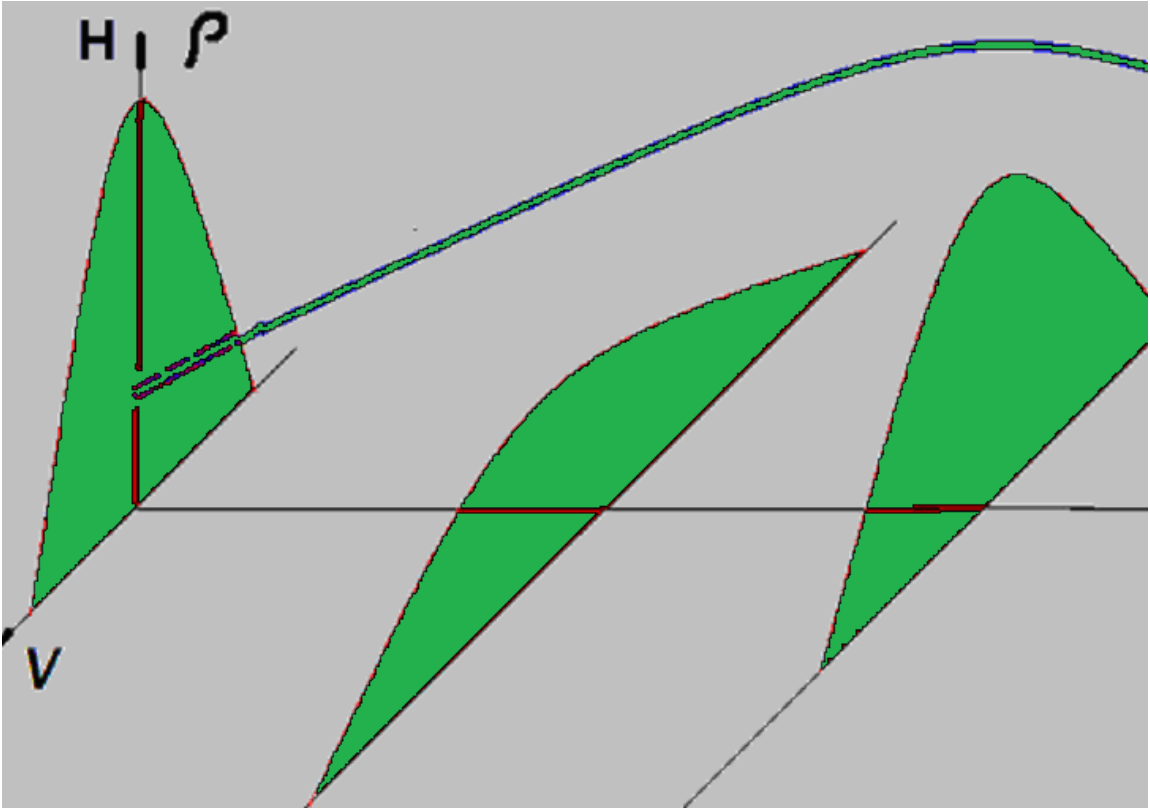

**Figure 4. Living system: Deviation from classical thermodynamics.**

Obviously the last statement should not be taken literary; indeed, the proposed model captures only those aspects of the living systems that are associated with their ***behavior***, and in particular, with their motor-mental dynamics, since other properties are beyond the dynamical formalism.  Therefore, such physiological processes that are needed for the metabolism, reproduction, est., are not included into the model. That is why this model is in a formal disagreement with the first and second laws of thermodynamics while the living systems are not. Indeed, applying the first law of thermodynamics we imply violation of conservation of ***mechanical*** energy since other types of energies (chemical, electro-magnetic, etc) are beyond our mathematical formalism. Applying the second law of thermodynamics, we consider our system as isolated one while the underlying real system is open due to other activities of livings that were not included in our model. Nevertheless, despite these limitations, the proposed model captures the "magic" of Life: the ability to create self-image and self-awareness, and that fits perfectly to the concept of ***intelligent agent.*** Actually the proposed model represents governing equations for interactions of intelligent agents. In order to emphasize the autonomy of the agents' decision making process, we will associate the proposed models with ***self-supervised (SS) active systems.***

By an active system we will understand here a set of interacting intelligent agents capable of processing information, while an intelligent agent is an autonomous entity which observes and acts upon an environment and directs its activity towards achieving goals. The active system is not derivable from the Lagrange or Hamilton principles, but it is rather created for information processing. One of specific differences between active and physical systems is that the former are supposed to act in uncertainties originated from incompleteness of information. Indeed, an intelligent agent almost never has access to the whole truth of its environment. Uncertainty can also arise because of incompleteness and incorrectness in the agent's understanding of the properties of the environment. That is why *quantum-inspired SS* systems are well suited for representation of active systems.

***f. Self-supervised active systems with integral feedback***. In this sub-section we will introduce a feedback from the mental to motor dynamics that is different from the feedback (6) discussed above. This feedback will make easier to formulate new principals of the game theory.

Let us introduce the following feedback, [8]

$$f = \frac{1}{\rho(v,t)} \int_{-\infty}^{v} [\rho(\zeta,t) - \rho^*(\zeta)] dv \qquad (49)$$

With the feedback (49), Eqs. (7) and (8) take the form, respectively

$$\dot{v} = \frac{1}{\rho(v,t)} \int_{-\infty}^{v} [\rho(\zeta,t) - \rho^*(\zeta) \qquad (50)$$

$$\frac{\partial \rho}{\partial t} + \rho(t) - \rho^* = 0 \qquad (51)$$

The last equation has the analytical solution

$$\rho = (\rho_0 - \rho^*)e^{-t} + \rho^* \qquad (52)$$

Subject to the initial condition

$$\rho(t=0) = \rho_0 \qquad (53)$$

This solution converges to a prescribed, or target, stationary distribution $\rho^*(V)$. Obviously the normalization condition for $\rho$ is satisfied if it is satisfied for $\rho_0$ and $\rho^*$. This means that Eq. (51) has an attractor in the probability space, and this attractor is stochastic. Substituting the solution (52) in to Eq. (50), one arrives at the ODE that simulates the stochastic process with the probability distribution (52)

$$\dot{v} = \frac{e^{-t}}{[\rho_0(v) - \rho^*(v)]e^{-t} + \rho^*(v)} \int_{-\infty}^{v} [\rho_0(\zeta) - \rho^*(\zeta)] \, d\zeta \qquad (54)$$

As notices above, the randomness of the solution to Eq. (54) is caused by instability that is controlled by the corresponding Liouville equation. It should be emphasized that in order to run the stochastic process started with the initial distribution $\rho_0$ and approaching a stationary process with the distribution $\rho^*$, one should substitute into Eq. (54) the *analytical expressions* for these functions.

It is reasonable to assume that the solution (4) starts with sharp initial condition

$$\rho_0(V) = \delta(V) \qquad (55)$$

As a result of that assumption, all the randomness is supposed to be generated *only* by the controlled instability of Eq. (54). Substitution of Eq. (55) into Eq. (54) leads to two different domains of $v$: $v \neq 0$ and $v=0$ where the solution has two different forms, respectively

$$\int_{-\infty}^{v} \rho^*(\zeta) d\zeta = \frac{C_1}{e^{-t} - 1}, \qquad v \neq 0 \qquad (56)$$

$$v \equiv 0 \qquad (57)$$

Eq. (57) represents a singular solution, while Eq. (56) is a regular solution that include arbitrary constant $C$. The regular solutions is unstable at $t=0$, $|v| \to 0$ where the Lipschitz condition is violated

$$\frac{d\dot{v}}{dv} \to \infty \, at \qquad t \to 0 \quad ,| v| \to 0 \qquad (58)$$

and therefore, an initial error always grows generating *randomness*.

Let us analyze the behavior of the solution (56) in more details. As follows from this solution, all the particular solutions intersect at the same point $v=0$ at $t=0$, and that leads to non-uniqueness of the solution due to violation of the Lipschitz condition. Therefore, the same initial condition $v=0$ at $t=0$ yields infinite number of different solutions forming a family (56); each solution of this family appears with a certain probability guided by the corresponding Liouville equation (51). For instance, in cases plotted in Fig.5 and Fig.6, the "winner" solution is, respectively,

$$v_1 = \varepsilon \to 0, \qquad \rho(v_1) = \rho_{max}, \qquad and \qquad v = v_2, \quad \rho(v_2) = \sup\{\rho\} \qquad (59)$$

since it passes through the maximum of the probability density (51). However, with lower probabilities, other solutions of the family (53) can appear as well. Obviously, this is a non-classical effect. Qualitatively, this property is similar to those of quantum mechanics: the system keeps all the solutions simultaneously and displays each of them "by a chance", while that chance is controlled by the evolution of probability density (51).

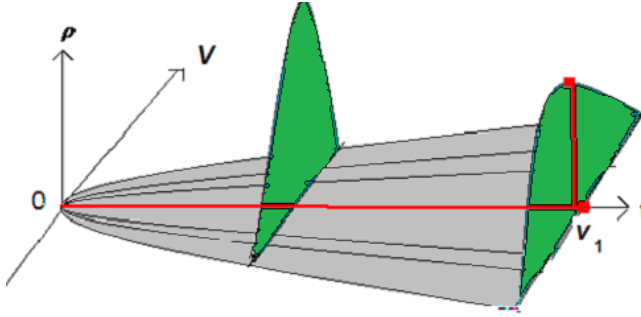 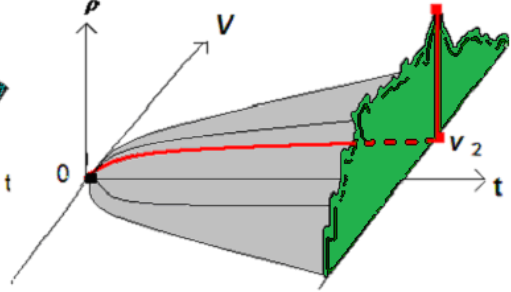

**Figure 5    Stochastic process and probability density**

**Figure 6    Global maximum**

The approach is generalized to *n*-dimensional case simply by replacing $v$ with a vector $v = v_1, v_2, \dots v_n$ since Eq. (51) does not include space derivatives.

*Examples.* Let us start with the following normal distribution

$$\rho^*(V) = \frac{1}{\sqrt{2\pi}} e^{-\frac{V^2}{2}} \tag{60}$$

Substituting the expression (60) and (55) into Eq. (56) at $V=v$, one obtains

$$v = erf^{-1}(\frac{C_1}{e^{-t}-1}), \qquad v \neq 0 \tag{61}$$

As another example, let us choose the target density $\rho^*$ as the Student's distribution, or so called power law distribution

$$\rho^*(V) = \frac{\Gamma(\frac{v+1}{2})}{\sqrt{v\pi}\,\Gamma(\frac{v}{2})} (1+\frac{V^2}{v})^{-(v+1)/2} \tag{62}$$

Substituting the expression (62) into Eq. (56) at $V=v$, and $v=1$, one obtains

$$v = \tan(\frac{C}{e^{-t}-1}) \quad for \qquad v \neq 0 \tag{63}$$

**g. Finding global maximum.** Based upon the proposed model with integral feedback, a simple algorithm for finding a global maximum of an *n*-dimensial function can be formulated. The idea of the proposed algorithm is very simple: based upon the model with integral feedback (50), and (51),  introduce  a *positive* function $\psi(v_1, v_2, \dots v_n)$, $|v_i| < \infty$ to be maximized as the probability density $\rho^*(v_1, v_2, \dots v_n)$ to which the solution of Eq. (50) is attracted. Then the larger value of this function will have the higher probability to appear. The following steps are needed to implement this algorithm.

1. Build and implement the *n*-dimensional version of the model Eqs. (50), and (51), as an analog devise

$$\dot{v}_i = \frac{e^{-t}}{n\{[\rho_0(v)-\rho^*(v)]e^{-t}+\rho^*(v)\}} \int_{-\infty}^{v_i} [\rho_0(\zeta) - \rho^*(\zeta)] \, d\zeta \qquad, \qquad i = 1, 2, \dots n. \tag{64}$$

2. Normalize the function to be maximized

$$\overline{\psi}(\{v\}) = \frac{\psi(\{v\})}{\int_{-\infty}^{\infty}\psi(\{v\})d\{v\}} \tag{65}$$

3. Using Eq. (51), evaluate time $\tau$ of approaching the stationary process to accuracy $\varepsilon$

$$\tau \approx \ln\frac{1-\overline{\psi}}{\varepsilon\overline{\psi}} \tag{66}$$

4. Substitute $\overline{\psi}$ instead of $\rho^*$ into Eqs. (64) and run the system during the time interval $\tau$.

5. The solution will "collapse" into one of possible solutions with the probability $\overline{\psi}$ .Observing (measuring) the corresponding values of $\{v*\}$, find the first approximation to the optimal solution.

6. Switching the device to the initial state and then starting again, arrive at the next approximations.

7. The sequence of the approximations represents Bernoulli trials that exponentially improve the chances of the optimal solution to become a winner. Indeed, the probability of success $\rho_s$ and failure $\rho_f$ after the first trial are, respectively

$$\rho_s = \overline{\psi}_1, \ \rho_f = 1-\overline{\psi}_1 \tag{67}$$

Then the probability of success after M trials is

$$\rho_{sM} = 1-(1-\overline{\psi})^M \to 1 \qquad at \qquad M \to \infty \tag{68}$$

Therefore, after *polinomial* number of trials, one arrived at the solution to the problem (unless the function $\psi$ is flat).

The main advantage of the proposed methodology is in a weak restriction imposed upon the *space structure* of the function $\overline{\psi}(\{x\})$ : it should be only integrable since there is no space derivatives included in the model (64). This means that $\overline{\psi}(\{v\})$ is not necessarily to be differentiable. For instance, it can be represented by a Weierstrass-like function $f(x) = \sum_{0}^{\infty} a^n \cos(b^n \pi x)$ , where $0 < a < 1$, $b$ is a positive odd integer, and $ab > 1+1.5\pi$ .

In a particular case when $\overline{\psi}(\{x\})$ is twice-differentiable, the algorithm is insensitive to local maxima because it is driven *not by gradients*, but by the *values* of this function.

***h. Entanglement in SS active systems with integral feedback***. We will continue the analysis of the SS system with integral feedback introduced above proceeding with the two-dimensional case

$$\dot{v}_1 = \frac{e^{-t}}{2[\rho_0(v_1,v_2)-\rho^*(v_1,v_2)]e^{-t}+\rho^*(v_1,v_2)}\int_{-\infty}^{v_1}[\rho_0(\zeta,v_2)-\rho^*(\zeta,v_2)]\ d\zeta \tag{69}$$

$$\dot{v}_2 = \frac{e^{-t}}{2[\rho_0(v_1,v_2)-\rho^*(v_1,v_2)]e^{-t}+\rho^*(v_1,v_2)}\int_{-\infty}^{v_2}[\rho_0(v_1,\zeta)-\rho^*(v_1,\zeta)]\ d\zeta \tag{70}$$

From these equations one obtains

$$\frac{dv_2}{dv_1} = \frac{F_2(v_1,v_2)}{F_1(v_1,v_2)} \tag{71}$$

where $F_1 = \int_{-\infty}^{v_1}[\rho_0(\zeta,v_2)-\rho^*(\zeta,v_2)]\ d\zeta$ , $\quad F_2 = \int_{-\infty}^{v_2}[\rho_0(v_1,\zeta)-\rho^*(v_1,\zeta)]\ d\zeta$ (72)

It should be recalled that the integrands, and therefore, $F_1,F_2$ in Eq. (72) are *known* functions of the state variables $v_1,v_2$ . For small deviations of the state variables from zero, Eq. (71) can be linearized as following

$$\frac{dv_2}{dv_1} = \frac{b_{21}v_1 + b_{22}v_2}{b_{11}v_1 + b_{12}v_2} \quad \text{where} \quad b_{ij} = \frac{\partial F_i}{\partial v_j}\big|_{v_i, v_j = +0} \tag{73}$$

This equation has the same structure as Eq. (43), and therefore,

$$v_2 = \widetilde{C}v_1, \qquad \widetilde{C} = const. \tag{74}$$

where both $v_1$ and $v_2$ are random.

The $n$-dimensional case of the $SS$ system can be written as

$$\dot{v}_i = \frac{1}{n\rho(v_1 \ldots v_n, t)} \int_{-\infty}^{v_i} [\rho(\zeta_i, v_{j \neq i}, t) - \rho^*(\zeta_i, v_{j \neq i})] d\zeta_i \quad i = 1, 2 \ldots n \tag{75}$$

where

$$\rho = (\rho_0 - \rho^*)e^{-t} + \rho^* \tag{76}$$

### 3. Application to game theory.

In this section we will address a situation when agents are competing. That means that they have *different* objectives. Turning to Eq.(75), (76), one can rewrite them for the case of competing agents

$$\dot{v}_i = \frac{1}{n\rho(v_{1,} \ldots v_n, t)} \int_{-\infty}^{v_i} [\rho(\zeta_i, v_{j \neq i}, t) - \rho_i^*(\zeta_i, v_{j \neq i})] d\zeta_i \tag{77}$$

$$\frac{d\rho}{dt} + \rho(t) - \frac{1}{n}\sum_{k=1}^{n} a_k \rho_k^* = 0 , \quad \sum_{k=1}^{n} a_k = n \qquad \rho_i^* \neq \rho_j^* \text{ if } i \neq j. \tag{78}$$

where $\rho_k^*$ is the preset density of the $k^{th}$ agent that can be considered as his objective, $a_k$ is a constant weight of the $k^{th}$ agent's effort to approach his objective.

Thus, each $k^{th}$ agent is trying to establish his own static attractor $\rho_k^*$, but due to entanglement, the whole system will approach the weighted average

$$\rho = [\rho^0 - \frac{1}{n}\sum_{i=1}^{n}(a_i \rho_i^*)]e^{-t} + \frac{1}{n}\sum_{i=1}^{n} a_i \rho_i^* \tag{79}$$

$$\rho(t) = \frac{1}{n}\sum_{1}^{n} a_k \rho_k^* \quad \text{at} \quad t \to \infty \tag{80}$$

Substituting the solution (79) into Eqs. (80), one arrives at a coupled system of $n$ ODE with respect to $n$ state variables $v_i$. Although a closed form analytical solution of the system (77) and (78) is not available, its property of the Lipcshitz instability at $t=0$ could be verified. This means that the solution to the system (77) and (78) is random, and if the system is run many times, the statistical properties of the whole ensemble will be described by Eq. (79). Obviously, those agents who have chosen density with a sharp maximum are playing more risky game. Here we have assumed that competeng agents are still entangled, and therefore, their information about each other is complete. More complex situation when the agents are not entangled, and exchanged information is incomplete is address in the next section. The simplest way to formalize the incompleteness of information possessed by competing agent is to include the "vortex" terms into Eqs. (77): these terms could change each particular trajectory of the agent motion, but they would not change the statistical invariants that remain available to the competing agents

$$\dot{v}_i = \frac{1}{n\rho(v_1 \ldots v_n, t)} [\int_{-\infty}^{v_i} [\rho(\zeta_i, v_{j \neq i}, t) - \rho^*(\zeta_i, v_{j \neq i})] d\zeta_i + \sum_{j \neq i}^{n} T_{ij} \tanh v_j]$$

It is easily verifiable that the augmented neural-net-like terms do not effect the corresponding Liouville equation, and therefore, they do not change the static attractor in the probability space described by Eq. (76). However, they may significantly change the configuration of the random trajectories in physical space making their entanglement more sophisticated. Another way to formalize uncertainty is to introduce a complex joint probability density where its imaginary part represents a measure of uncertainty in density distribution. This case will be considered below in more details.

***a. Problem formulation***. In this section we will present a draft of application of self-supervised *active* dynamical systems to differential games. Following von-Neuman, and Isaaks, [1], we will introduce a two-player zero-sum (antagonistic) differential game that is described by dynamical equations (64)

rewritten for *i=1,2*

$$\dot{v}_i = \frac{1}{2\rho(v_1,...v_n,t)} \int_{-\infty}^{v_i} [\rho(\zeta_i, v_{j\neq i}, t) - a_i C_i f_i(\zeta_i, v_{j\neq i})] d\zeta_i \qquad f_1 = f, \quad f_2 = f^{-1}, \quad a_1 + a_2 = 2. \quad (82)$$

where $v$ is the state vector, $f_1$ is the control vector of the maximizing player $E$ (evader), $f_2$ is the control vector of the minimizing player $P$ (pursuer), and $C_1, C_2$ are the normalizing factors. Obviously, $f$ is a known function of both state variables.

However, the rules of the game we propose is slightly different from those introduced by Isaaks, namely: the player $P$ tries to minimize the function $f$, (i.e. maximize the function $f^{-1}$) while the player $E$ tries to maximize $f$ in the same manner as it was described in the previous sub-section i.e. via entanglement. The Liouville equation for the system (69) follows from Eq. (78 )

$$\frac{d\rho}{dt} + \rho(t) - \frac{1}{2}(a_1 C_1 f + a_2 C_2 f^{-1}) = 0 \qquad (83)$$

whence

$$\rho = [\rho^0 - \frac{1}{2}(a_1 C_1 f + a_2 C_2 f^{-1})]e^{-t} + \frac{1}{2}(a_1 C_1 f + a_2 C_2 f^{-1}), \qquad (84)$$

We will now give a description of the game.
The game starts with zero initial conditions:

$$v_1 = 0, \quad v_2 = 0, \quad \rho^0 = \delta(v_1, v_2) \qquad at \qquad t = 0 \qquad (85)$$

It is assumed that each player has access to the system (82), (84), and therefore, he has complete information about its state. The substitution of Eq. (84) into Eqs. (82) closes the system (82). However, because of a failure of the Lipcshitz condition at *t=0* (see Eq. (58)), the solution of Eqs. (82) is random, and each player can predict it only in terms of probability. As follows from Eq. (84), the highest probability to appear has the solution that delivers the global maximum to the payoff function

$$F = a_1 C_1 f + a_2 C_2 f^{-1} \qquad (86)$$

Obviously, the player that has higher weight $a_i$ would have better chances to win since the global maximum of Eq. (86) is closer to the global maximum of his goal function. With reference to Eq.(86) , a player can evaluate time $\tau$ of approaching the stationary process to accuracy $\varepsilon$ as

$$\tau \approx \ln\frac{1-F}{\varepsilon F} \qquad (87)$$

and introduce

$$v_1^{\ 1} = v_1(\tau), \qquad v_2^{\ 1}(\tau), \quad f_1^{\ 1} = f_1(v_1^{\ 1}, v_2^{\ 2}), \quad f_2^{\ 1} = f_2(v_1^{\ 1}, v_2^{\ 1}) \qquad (88)$$

This is the end of the first move. After that, each player updates his weight as following

$$a_1^{\ 1} = a_1 + \frac{f_1^{\ 1} - f_1^{\ 0}}{f_1^{\ 0}}, \qquad a_2^{\ 1} = a_2 - \frac{f_1^{\ 1} - f_1^{\ 0}}{f_1^{\ 0}}, \quad f_1^{\ 0} \neq 0, \qquad (89)$$

and starts the next move with the same initial conditions (85). But the system (82), (84) is different now: the control functions $f_1, f_2$ are to be replaced by their updated values $f_1^{\ 1}, f_2^{\ 1}$ respectively. Thus, during the first move, the potential winner is selected by a chance, and during the next move, his chances are increased due to favorable update of the weights. However, the role of the chance is still significant even during the subsequent moves; indeed, if the global maximum of the control function $F$ is sharp, the initially selected potential winner still can lose.
The game ends when one of the players achieves his goal my maximizing his control function to a preset level, for instance, if

$$f_1 - f_2 > A^2 \qquad (90)$$

***b. Games with incomplete information***. The theory presented above includes applications to such problems as battle games, games with moving craft, pursuit games, etc. However, the main limitation of this theory, as well as the most of the game theories, is that it requires complete information about the state variables available to both players. This limitation significantly diminishes the applicability of the theory to real-life games where the complete information is not available. That is why the extension of this theory to cases of incomplete information is of vital importance. In our application, we will assume that each player knows only his own state variables, while he has to guess about the state variables of his adversary. For that case, the mathematical formalism of *SS* systems can offer a convenient tool to replace unknown value of a state variable by its expected value. Such a possibility is available due to players' dependence (but not necessarily entanglement) via the joint probability density: since each player possesses the joint density, he can, at any moment, compute the expected value of the state variable of the other player.

We assume that the players follow the strategy: "what do you think I think you think…?" and we will start with the assumption that each player takes a conservative view by thinking that although he does not know the values of the state variable of his adversary, the adversary *does know* the values of his state variable. Then the governing equation for the Evader will be

$$\dot{v}_1 = \frac{1}{2\rho(v_1, \overline{V}_2, t)} \int_{-\infty}^{v_1} \rho(\zeta_1, \overline{V}_2, t) - a_1 C_1 f_1(\zeta_1, \overline{V}_2) d\zeta_1 \tag{91}$$

Here $\overline{V}_2$ is the expected value of $V_2$

$$\overline{V}_2 = \int_{-\infty}^{\infty} V_2 \rho(V_1, V_2) dV_1 dV_2 \tag{92}$$

Now the Evader has to create the *image* of the Pursuer by using the expected value of his state variable

$$\dot{v}_{2|1} = \frac{1}{2\rho(v_1, \overline{V}_2, t)} \int_{-\infty}^{v_2} [\rho(v_1, \zeta_2, t) - a_2 C_2 f_2(v_1, \zeta_2)] d\zeta_2 \tag{93}$$

where $v_{2|1}$ is the state variable of the Pursuer in view of the Evader.

The corresponding Liouville equation that governs the joint probability equation is not changed: it is still given by Eq. (78). Its solution (84) should be substituted into Eqs. (91) and (93) along with the Eq. (92). Obviously, the expected value (92) is found from the solution (84)

$$\overline{V}_i = [\overline{V}_i^0 - \int_{-\infty}^{\infty} d\zeta_1 \int_{-\infty}^{\infty} d\zeta_2 \{[\frac{1}{2}(a_1 C_1 f + a_2 C_2 f^{-1})]e^{-t} + \frac{1}{2}(a_1 C_1 f + a_2 C_2 f^{-1})]\overline{V}_i\}, \quad i = 1, 2. \tag{94}$$

The system of Eqs, (91), (93), and (78) with reference to Eqs. (92), (94) is closed.

Similar system can be obtained for governing equation of the Pursuer coupled with the governing equation of his *image* of the Evador:

$$\dot{v}_2 = \frac{1}{2\rho(\overline{V}_1, v_2, t)} \int_{-\infty}^{v_1} \rho(\overline{V}_1, \zeta_2, t) - a_2 C_2 f_2(\overline{V}_1, \zeta_2) d\zeta_2 \tag{95}$$

$$\dot{v}_{1|2} = \frac{1}{2\rho(\overline{V}_1, v_2, t)} \int_{-\infty}^{v_1} \rho(\zeta_1, v_2, t) - a_2 C_2 f_2(\zeta_1, v_2) d\zeta_1 \tag{96}$$

$$\overline{V}_1 = \int_{-\infty}^{\infty} V_1 \rho(V_1, V_2) dV_1 dV_2 \tag{97}$$

After substitution Eqs. (78) into Eqs, (95) and (96), with reference to Eqs. (97) and (94), one arrives at the closed system.

Thus we obtain two independent systems of ODE describing *entanglement* of the player with the image of his adversary. Each system has random solutions that appear with the probability described by Eq. (78). After time interval $\tau$ (see Eq. (87)), each player gets access to the real values of the functions $f_i$ to be maximized, and based upon that, he can update the state variables and weights for the next move, (see Eqs. (88) and (89)).

Let us consider now the case when the players do not know the values of state variables of their adversary. Then instead of the systems (91),(93) and (95),(96) we have, respectively

$$\dot{v}_1 = \frac{1}{2\rho(v_1, \overline{V}_2, t)} \int_{-\infty}^{v_1} \rho(\zeta_1, \overline{V}_2, t) - a_1 C_1 f_1(\zeta_1, \overline{V}_2) d\zeta_1 \tag{98}$$

$$\dot{v}_{2|1|1} = \frac{1}{2\rho(\overline{V}_1, \overline{V}_2, t)} \int_{-\infty}^{v_2} [\rho(\overline{V}_1, \zeta_2, t) - a_2 C_2 f_2(\overline{V}_1, \zeta_2)] d\zeta_2 \tag{99}$$

$$\dot{v}_2 = \frac{1}{2\rho(\overline{V}_1, v_2, t)} \int_{-\infty}^{v_1} \rho(\overline{V}_1, \zeta_2, t) - a_2 C_2 f_2(\overline{V}_1, \zeta_2) d\zeta_2 \tag{100}$$

$$\dot{v}_{1|2|2} = \frac{1}{2\rho(\overline{V}_1, \overline{V}_2, t)} \int_{-\infty}^{v_1} \rho(\zeta_1, \overline{V}_2, t) - a_2 C_2 f_2(\zeta_1, \overline{V}_2) d\zeta_1 \tag{101}$$

Here $v_{2|1|1}$ is the state variable of the Pursuer's view on the Evader in view of the Pursuer, and $v_{1|2|2}$ is the state variable of the Evader's view on the Pursuer in view of the Evader.

It is easy to conclude that the image equations (99) and (101) can be solved independently

$$v_{2|1|1} = \int_0^t dt \frac{1}{2\rho(\overline{V}_1, \overline{V}_2, t)} \int_{-\infty}^{v_2} [\rho(\overline{V}_1, \zeta_2, t) - a_2 C_2 f_2(\overline{V}_1, \zeta_2)] d\zeta_2 \tag{102}$$

$$v_{1|2|2} = \int_0^t dt \frac{1}{2\rho(\overline{V}_1, \overline{V}_2, t)} \int_{-\infty}^{v_1} \rho(\zeta_1, \overline{V}_2, t) - a_2 C_2 f_2(\zeta_1, \overline{V}_2) d\zeta_1 \tag{103}$$

Now replacing $\overline{V}_2, \overline{V}_1$ in Eqs.(98) and (100) by the solutions for $v_{2|1|1}$ and $v_{1|2|2}$, respectively, one arrives at two independent ODE describing behaviors of the players. Therefore, at this level of incompleteness of information, the entanglement disappears.

The games with incomplete information give a reason to distinguish two type of dependence between the agents described by the variables $v_i$ in the $iQ$ systems. The first type of dependence is entanglement that has been introduced and discussed above. One should recall that in order to be entangled, the agents are supposed to run the system jointly during some initial period of time. But what happens if the agents had never been in contact? Obviously they are not entangled, i.e. they cannot predict each other motions. However they are not completely independent: they can make random decisions, but the probability of these decisions will be correlated via the joint probability. As a result, the agents will be able to predict expected decisions of each other. We will call such correlation a ***weak*** entanglement. As follows from the games with incomplete information considered above, weak entanglement was presented as entanglement of an agent with the probabilistic image of another agent.

**4. Games of partially entangled agents.**

In this section we introduce a new, more sophisticated entanglement that does not exists in quantum mechanics, but can be found in SS models. This finding is based upon existence of incompatible stochastic processes that are considered below.

***a. Incompatible stochastic processes.*** Classical probability theory defines conditional probability densities based upon the existence of a joint probability density. However, one can construct correlated stochastic processes that are represented only by conditional densities since a joint probability density does not exist. For that purpose, consider two coupled Langevin equations [9]

$$\dot{x}_1 = g_{11}(x_2) L_1(t) \tag{104}$$

$$\dot{x}_2 = g_{22}(x_1) L_2(t) \tag{105}$$

where the Langevin forces $L_1(t)$ and $L_2(t)$ satisfy the conditions

$$\langle L_i(t) \rangle = 0, \qquad \langle L_i(t) L_i(t') \rangle = 2g_{ii}\delta(t - t') \tag{106}$$

Then the joint probability density $\rho(X_1, X_2)$ describing uncertainties in values of the random variables $x_1$ and $x_2$ evolves according to the following Fokker-Planck equation

$$\frac{\partial \rho}{\partial t} = g_{11}^{\ 2}(X_2)\frac{\partial^2 \rho}{\partial X_1^{\ 2}} + g^2_{\ 22}(X_1)\frac{\partial^2 \rho}{\partial X_2^{\ 2}} \qquad (107)$$

Let us now modify Eqs. (104) and (105) as following

$$\dot{x}_1 = g_{11}^{\ 2}(x^*_2)L_1(t) \qquad (108)$$

$$\dot{x}_2 = g_{22}^{\ 2}(x^*_1)L_2(t) \qquad (109)$$

where $x_1^*$ and $x_2^*$ are fixed values of $x_1$ and $x_2$ that play role of parameters in Eqs. (108) and (109), respectively. Now the uncertainties of $x_1$ and $x_2$ are characterized by conditional probability densities $\rho_1(X_1 \mid X_2)$ and $\rho_2(X_2 \mid X_1)$ while each of these densities is governed by its own Fokker-Planck equation

$$\frac{\partial \rho_1}{\partial t} = g_{11}^{\ 2}(X_2)\frac{\partial^2 \rho_1}{\partial X^2_{\ 1}} \qquad (110)$$

$$\frac{\partial \rho_2}{\partial t} = g_{22}^{\ 2}(X_1)\frac{\partial^2 \rho_2}{\partial X_2} \qquad (111)$$

The solutions of these equations subject to sharp initial conditions

$$\rho_i(X_i, t \mid X_i', t') = \delta(X_i - X_i'), \qquad i = 1, 2. \qquad (112)$$

for $t > t'$ read

$$\rho_1(X_1 \mid X_2) = \frac{1}{\sqrt{4\pi g_{11}^{\ 2}(X_2)(t-t')}}\exp(-\frac{(X_1 - X_1')^2}{4g_{11}^{\ 2}(X_2)(t-t')} \qquad (113)$$

$$\rho_2(X_2 \mid X_1) = \frac{1}{\sqrt{4\pi g_{22}^{\ 2}(X_1)(t-t')}}\exp(-\frac{(X_2 - X_2')^2}{4g^2_{\ 22}(X_1)(t-t')} \qquad (114)$$

As shown in [5], a joint density for the conditional densities (113) and (114) exists only in special cases of the diffusion coefficients $g_{11}$ and $g_{22}$ when the conditional probabilities are compatible. These conditions are

$$ink(\rho_1, \rho_2) = \frac{\partial}{\partial X_1 \partial X_2}\ln\frac{\rho_1(X_1 \mid X_2)}{\rho_2(X_2 \mid X_1)} \equiv 0 \qquad (115)$$

Indeed

$$\rho(X_1, X_2) = \rho_1(X_1 \mid X_2)\int_{-\infty}^{\infty}\rho(\zeta, X_2)d\zeta = \rho_2(X_2 \mid X_1)\int_{-\infty}^{\infty}\rho(X_1, \zeta)d\zeta, \qquad (116)$$

whence

$$\ln\frac{\rho_1(X_1 \mid X_2)}{\rho_2(X_2 \mid X_1)} = \ln\int_{-\infty}^{\infty}\rho(X_1, \zeta)d\zeta - \ln\int_{-\infty}^{\infty}\rho(\zeta, X_2)d\zeta \qquad (117)$$

and that leads to Eq. (115).

Thus, the existence of the join density $\rho(X_1, X_2)$ for the conditional densities $\rho_1(X_1 \mid X_2)$ and $\rho_2(X_2 \mid X_1)$ requires that

$$\frac{\partial^2}{\partial X_1 \partial X_2}[\frac{(X_1 - X_1')^2}{4g_{11}^{\ 2}(X_2)} - \frac{(X_2 - X_2')^2}{4g_{22}^{\ 2}(X_1)}] \equiv 0 \qquad (118)$$

Obviously the identity (118) holds only for specially selected functions $g_{11}(X_2)$ and $g_{22}(X_1)$, and therefore, existence of the joint density is an exception rather than a rule.

**b. Partial Entanglement.** In order to prove existence of a new form of entanglement, let us modify the system Eqs. (36), (37), and (38) as following

$$\dot{v}_1 = -a_{11}(v_2)\frac{\partial}{\partial v_1}\ln\rho_1(v_1\mid v_2) \tag{119}$$

$$\frac{\partial\rho_1(V_1\mid V_2)}{\partial t} = a_{11}(V_2)\frac{\partial^2\rho_1(V_1\mid V_2)}{\partial V^2_1} \tag{120}$$

$$\dot{v}_2 = -a_{22}(v_1)\frac{\partial}{\partial v_1}\ln\rho_2(v_2\mid v_1) \tag{121}$$

$$\frac{\partial\rho_2(V_2\mid V_1)}{\partial t} = a_{22}(V_1)\frac{\partial^2\rho_2(V_2\mid V_1)}{\partial V^2_2} \tag{122}$$

Since here we do not postulate existence of a joint density, the system is written in terms of conditional densities, while Eqs. (120) and (121) are similar to Eqs. (110) and (111). The solutions of these PDE can be written in the form similar to the solutions (113) and (114)

$$\rho_1(V_1\mid V_2) = \frac{1}{\sqrt{4\pi a_{11}(V_2)(t-t')}}\exp(-\frac{(V_1-V_1')^2}{4a_{11}(V_2)(t-t')} \tag{123}$$

$$\rho_2(V_2\mid V_1) = \frac{1}{\sqrt{4\pi a_{22}(V_1)(t-t')}}\exp(-\frac{(V_2-V_2')^2}{4a_{11}(V_1)(t-t')} \tag{124}$$

As noticed in the previous sub-section, the existence of the joint density $\rho(V_1,V_2)$ for the conditional densities $\rho_1(V_1\mid V_2)$ and $\rho_2(V_2\mid V_1)$ requires that

$$\frac{\partial^2}{\partial V_1\partial V_2}[\frac{(V_1-V_1')^2}{4a_{11}(V_2)} - \frac{(V_2-V_2')^2}{4a_{22}(V_1)}]\equiv 0 \tag{125}$$

In this case, the joint density exists (although its finding is not trivial, [9]), and the system (119)-(122) can be reduced to a system similar to (36)-(38). But here we will be interested in case when the joint density does **not** exist. It is much easier to find such functions $a_{11}(V_2), a_{22}(V_1)$ for which the identity (125) does **not** hold, and we assume that

$$\frac{\partial^2}{\partial V_1\partial V_2}[\frac{(V_1-V_1')^2}{4a_{11}(V_2)} - \frac{(V_2-V_2')^2}{4a_{22}(V_1)}]\neq 0 \tag{126}$$

In this case the system (119)-(122) cannot be simplified. In order to analyze this system in details, let substitute the solutions (123) and (124) into Eqs.(119) and (121), respectively. Then with reference to Eq. (14), one obtains

$$\dot{v}_1 = \frac{v_1}{2t} \tag{127}$$

$$\dot{v}_2 = \frac{v_2}{2t} \tag{128}$$

and therefore

$$v_1 = C_1\sqrt{t} \tag{129}$$

$$v_2 = C_2\sqrt{t} \tag{130}$$

It should be recalled that according to the terminology introduced in Section 3, the system (119)- (120) and the system (121)-(122) can be considered as dynamical models for interaction of two *communicating* agents where Eqs.(119) and (121) describes their motor dynamics , and Eqs. (120) and (122) – mental dynamics, respectively. Also it should be reminded that the solutions (129) and (130) are represented by one-parametrical families of random samples, as in Eq. (15), while the randomness enters through the time-independent parameters $C_1$ and $C_2$ that can take any real numbers. As follows from Fig.

2, all the particular solutions (129) and (130) intersect at the same point $v_{1,2} = 0$ at *t=0*, and that leads to non-uniqueness of the solution due to violation of the Lipcshitz condition. Therefore, the same initial condition $v_{1,2} = 0$ at *t=0* yields infinite number of different solutions forming a family; each solution of this family appears with a certain probability guided by the corresponding Fokker-Planck equations (120) and (122), respectively. Similar scenario was described in the Section 2 of this paper. But what unusual in the system (119)-(121) is correlations: although Eqs. (120) and (122) are correlated, and therefore, mental dynamics are entangled, Eqs (119) and (121) are *not* correlated (since they can be presented in the form of independent Eqs. (127) and (128), respectively), and therefore, the motor dynamics are *not* entangled. This means that in the course of communications, each agent "selects" a certain pattern of behavior from the family of solutions (129) and (130) respectively, and these patterns are independent; but the *probabilities* of these "selections" are entangled via Eqs. (120) and (122). Such sophisticated correlations *cannot* be found in physical world, and they obviously represent a "human touch". Unlike the entanglement in system with joint density (such as that in Eqs. (36)- (38)) here the agents do not share any deterministic invariants (compare to Eq. (48)). Instead the agents can communicate via "best guesses" based upon known conditional probability densities distributions.

In order to quantify the amount of uncertainty due to incompatibility of the conditional probability densities (123) and (124), let us introduce a concept of complex probability, [5],

$$f(V_1, V_2) = a(V_1, V_2) + ib(V_1, V_2) \qquad (131)$$

Then the marginal densities are

$$f_1(V_1) = \int_{-\infty}^{\infty} a(V_1, V_2)dV_2 + i\int_{-\infty}^{\infty} b(V_1, V_2)dV_2 = a_1(V_1) + ib_1(V_1) \qquad (132)$$

$$f_2(V_2) = \int_{-\infty}^{\infty} a(V_1, V_2)dV_1 + i\int_{-\infty}^{\infty} b(V_1, V_2)dV_1 = a_2(V_2) + ib_2(V_2) \qquad (133)$$

Following the formalism of conditional probabilities, the conditional densities will be defined as

$$f_{1|2} = \frac{f(V_1, V_2)}{f_2(V_2)} = \frac{a(V_1, V_2) + ib(V_1, V_2)}{a_2(V_2) + ib_2(V_2)} = \frac{aa_2 + bb_2}{a_2^2 + b_2^2} + i\frac{a_2b - ab_2}{a_2^2 + b_2^2} \qquad (134)$$

$$f_{2|1} = \frac{f(V_1, V_2)}{f_1(V_1)} = \frac{a(V_1, V_2) + ib(V_1, V_2)}{a_1(V_1) + ib_1(V_1)} = \frac{aa_1 + bb_1}{a_1^2 + b_1^2} + i\frac{a_1b - ab_1}{a_1^2 + b_1^2} \qquad (135)$$

with the normalization constraint

$$\int_{-\infty}^{\infty}\int_{-\infty}^{\infty} (a^2 + b^2)^{1/2} dV_1 dV_2 = 1 \qquad (136)$$

This constraint can be enforced by introducing a normalizing multiplier in Eq.131) which will not affect the conditional densities (134) and (135).

Clearly

$$a \le (a^2 + b^2)^{1/2}, \quad \text{and} \quad \int_{-\infty}^{\infty}\int_{-\infty}^{\infty} adV_1 dV_2 \le 1 \qquad (137)$$

Now our problem can be reformulated in the following manner: given two conditional probability densities (123) and (124), and considering them as real parts of (unknown) complex densities (134) and (135), find the corresponding complex joint density (131), and therefore, all the marginal (132) and (133), as well as the imaginary parts of the conditional densities. In this case one arrives at two coupled integral equations with respect to two unknowns $a(V_1, V_2)$ and $b(V_1, V_2)$ (while the formulations of $a_1(V_1, V_2)$, $a_2(V_1, V_2)$, $b_1(V_1, V_2)$ and $b_2(V_1, V_2)$ follow from Eqs.(132) and (133)). These equations are

$$\rho_1(V_1,V_2) = \frac{aa_2 + bb_2}{a_2^2 + b_2^2}, \qquad \rho_2(V_1,V_2) = \frac{aa_1 + bb_1}{a_2^2 + b_2^2}, \tag{138}$$

The system (138) is nonlinear, and very little can be said about general property of its solution without detailed analysis. Omitting such an analysis, let us start with a trivial case when

$$b = 0 \tag{139}$$

In this case the system (138) reduces to the following two integral equations with respect to *one* unknown

$$a(V_1, V_2)$$

$$\rho_1(V_1,V_2) = \frac{a(V_1,V_2)}{\displaystyle\int_{-\infty}^{\infty} a(V_1,V_2)dV_2}, \qquad\qquad \rho_2(V_1,V_2) = \frac{a(V_1,V_2)}{\displaystyle\int_{-\infty}^{\infty} a(V_1,V_2)dV_1}, \tag{140}$$

This system is overdetermined unless the compatibility conditions (115) are satisfied.

As known from classical mechanics, the incompatibility conditions are usually associated with a fundamentally new concept or a physical phenomenon. For instance, incompatibility of velocities in fluid (caused by non-existence of velocity potential) introduces vorticity in rotational flows, and incompatibility in strains describes continua with dislocations. In order to interpret the incompatibility (115), let us return to the system (138). Discretizing the functions in Eqs. (138) and replacing the integrals by the corresponding sums, one reduces Eqs. (138) to a system of *n* algebraic equations with respect to *n* unknowns. This means that the system is closed, and cases when a solution does not exist are exceptions rather than a rule. Therefore, in most cases, for any arbitrarily chosen conditional densities, for instant, for those given by Eqs, (123) and (124), the system (138) defines the complex joint density in the form (131).

Now we are ready to discuss a physical meaning of the imaginary component of the complex probability density. Firstly, as follows from comparison of Eqs. (138) and (140), the imaginary part of the probability density appears as a response to incompatibility of the conditional probabilities, and therefore, it can be considered as a "compensation" for the incompatibility. Secondly, as follows from the inequalities (137), the imaginary part consumes a portion of the "probability mass" increasing thereby the degree of uncertainty in the real part of the complex probability density. Hence the imaginary part of the probability density can be defined as a measure of the uncertainty "inflicted" by the incompatibility into the real part of this density.

In order to avoid solving the system of integral equations (138), we can reformulate the problem in an inverse fashion by assuming that the *complex* joint density is given. Then the real parts of the conditional probabilities that drive Eqs.(119) and (120) can be found from simple formulas (134) and (135).

Let us illustrate this new paradigm, and consider two players assuming that each player knows his own state but does not know the state of his adversary. In order to formalize the degree of initial incompleteness of information, introduce the complex joint probability density,

$$\rho_0(V_1,V_2) = a_0\delta(V_1,V_2) + ib_0\delta(V_1,V_2)$$

that shows how much the players know and how much they do not know about each other when the game starts. With reference to the normalization constraint (136),

$$(a_0^2 + b_0^2)^{1/2} = 1 \tag{142}$$

The structure of the real part of the joint probability density can be chosen the same as in Eq. (84)

$$\operatorname{Re}\rho = a_0\{[\delta - \frac{1}{2}(\alpha_1 C_1 f + \alpha_2 C_2 f^{-1})]e^{-t} + \frac{1}{2}(\alpha_1 C_1 f + \alpha_2 C_2 f^{-1})\}, \tag{143}$$

However since here $a_0 < 1$, the real part of the joint probability density is reduced due to a "leak" of the probability "mass" from the real to the imaginary part, and this makes predictions less certain for the both players. Otherwise the formal structure of the motor dynamics is similar to that described by Eqs. (83) and (84).

The imaginary part can be preset as

$$\operatorname{Im}\rho = b_0[(\delta - C_3\rho^*)e^{-t} + C_3\rho^*] \tag{144}$$

where $\rho^*$ is the probability density characterizing the degree of uncertainty of information that the players have about each other, while the larger $\rho^*$ the more the probability leak from the real to imaginary part of the complex probability density. The arbitrary constants $C_1$, $C_2$ and $C_3$ couples the real and the imaginary parts via the normalization constraint (136)

$$\int\limits_{-\infty}^{\infty}\int\limits_{-\infty}^{\infty}[(\operatorname{Re}\rho)^2+(\operatorname{Im}\rho^2)]^{1/2}\,dV_1dV_2=1 \tag{145}$$

The motor dynamics has a slight change compare to Eqs.(82)

$$\dot{v}_1=\operatorname{Re}\{\frac{e^{-t}}{2[\delta(v_1,v_2)-\rho^*(v_1,v_2)]e^{-t}+\rho^*(v_1,v_2)}\int\limits_{-\infty}^{v_1}[\delta(\zeta,v_2)-\rho^*(\zeta,v_2)]\,d\zeta\} \tag{146}$$

$$\dot{v}_2=\operatorname{Re}\{\frac{e^{-t}}{2[\delta(v_1,v_2)-\rho^*(v_1,v_2)]e^{-t}+\rho^*(v_1,v_2)}\int\limits_{-\infty}^{v_2}[\delta(v_1,\zeta)-\rho^*(v_1,\zeta)]\,d\zeta\} \tag{147}$$

Thus both players rely only upon the real part of the *complex* joint density instead of a *real* joint density (that may not exist in this case). But as follows from the inequalities (137), the values of density of the *real part* are lowered due to loss of the probability mass, and this increases the amount of uncertainty in player's predictions. In order to minimize that limitation, the players can invoke the imaginary part of the joint density that gives them *qualitative* information about the amount of uncertainty at the selected maxima.

It should be noticed that the game starts with a significant amount of uncertainties that will grow with next moves. Such subtle and sophisticated relationship is typical for communications between humans, and the proposed model captures it via partial entanglement introduced above.

Remark. So far we considered the imaginary part of a joint probability density as a result of incompatibility of conditional densities of the players. However this part can have a different origin: it can also represent a degree of deception that the players apply in real-life games. As in the previous example, in games with deception the imaginary part of the joint probability density increases uncertainty of the players' prediction capabilities. The mathematical formalism of the game with deception is similar to that discussed above.

## 6. Conclusion.

The paper combines several departures from classical methods in physics and in probability theory.

Firstly, it introduces a non-linear version of the Liouville equation that is coupled with the equation of motion (in Newtonian dynamics they are uncoupled). This new dynamical architecture grew up from quantum physics (in the Madelung version) when the quantum potential was replaced by information forces. The advantage of this replacement for modeling communications between intelligent agents representing living systems is addressed and discussed.

Secondly, it exploits a paradigm coming from incompatible conditional probabilities that leads to non-existence of a joint probability (in classical probability theory, existence of a joint probability is postulated). That led to discovery of a new type of entanglement that correlates not actions of Livings, but rather the probability of these actions.

Thirdly, it introduces a concept of imaginary probability as a measure of uncertainty generated by incompatibility of conditional probabilities.

All of these departures actually extend and complement the classical methods making them especially successful in analysis of communications in Living represented by new mathematical formalism.

Thus the paper introduces a fundamentally new approach to theory of differential games in which attention is concentrated upon behavioral properties of players as intelligent subjects possessing self-image and self-awareness. Due to quantum-like entanglement they are capable to predict and influence actions of their adversaries. The entanglement is generated by a joint probability density known by the players. In case of a complex density, its imaginary part represents a measure of uncertainty of the density distribution. The novelty of the approach is in non-Newtonian mathematical formalism that is based upon a ***behavioral*** model of Livings. The model is quantum inspired: it is represented by a modified Madelung equation in which the quantum potential is replaced by different, specially chosen "computational" potential. It consists of motor dynamics simulating actual behavior of the object, and mental dynamics representing evolution of the corresponding knowledge-base and incorporating this knowledge in the form of information flows into the motor dynamics. Due to feedback from mental dynamics, the motor dynamics attains quantum-like properties: its trajectory splits into a family of different trajectories, and each of those trajectories can be chosen with the probability prescribed by the mental dynamics. The model addresses a new type of entanglement that correlates the probabilities of actions of Livings rather than the actions themselves.

There are several differences between the proposed and conventional game theories. Firstly, in the proposed game, the players are entangled: they cannot make independent deterministic decisions; instead, they make coordinated random decisions such that, at least, the probabilities of these decisions are dependent. Therefore, the proposed game represents a special case of non-determine symmetric simultaneous zero-sum game. Secondly, the maximization of the payoff function here does not require any special methods (like gradient ascend) since it is "built-in" into the dynamical model. Indeed, the payoff function (86) is represented by the probability density of the stochastic attractor, and therefore, its maximum value has the highest probability to appear as a random solution of the underlying dynamical model (82). Moreover, the payoff function (86) is not required to be differentiable at all (although it must be integrable).

**References.**